\documentclass[11pt]{article}
\usepackage{moriond,epsfig}

\def\Journal#1#2#3#4{{#1} {\bf #2}, #3 (#4)}

\def\EPJ{{\em Eur. Phys. Jour.} C}

\def\mco{\multicolumn}

\def\be{\begin{equation}}
\def\ee{\end{equation}}
\def\bea{\begin{eqnarray}}
\def\eea{\end{eqnarray}}

\begin{document}
\vspace*{4cm}
\title{Early physics with top quarks at the LHC}

\author{ P. Ferrari\\ On behalf of the ATLAS and CMS Collaborations}

\address{CERN, \\
1211 Geneve 23, Switzerland}

\maketitle\abstracts{
The ATLAS and CMS experiments are now in their final installation phase
and will be soon ready to study the physics of proton-proton collisions 
at the Large Hadron Collider.
The LHC, by producing 2 $t\bar{t}$ events per second, 
will provide more than 8 million top events a year at start-up.
In this paper, particular emphasis is given to the $t\bar{t}$  
physics studies that can be performed at the beginning of the LHC running, with
a limited amount of integrated luminosity  ( $\le$10 fb$^{-1}$).}

\section{Introduction}

In the early days of data taking at the LHC, top physics will have a 
role of primary importance for several reasons.
First of all, since top physics allows for precise studies of the 
Standard Model (SM) and since the determination of 
the top mass  constraints the Higgs mass via radiative corrections.
At start-up, already with the first  few fb$^{-1}$ of integrated luminosity and with a non perfectly
calibrated detector, a top signal can be clearly separated from the background and the
top pair production cross-section can be extracted at better than 20\% accuracy and
with negligible statistical error.
The first measurement of the top mass will provide feedback on the detector
performance and top events can be used to understand and calibrate the detector light jet 
energy scale and the b-tagging. 
Additionally in scenarios beyond the SM, new particles may decay into top quarks,
therefore a detailed study of the top quark properties may provide a hint on new physics.
A good understanding of top physics is also essential since top events are a background 
for many new physics searches. 

\section{ Early selection of top events in the leptons+jets channel}

Since top events are so crucial for the initial phase of data taking, it is important to 
understand how much integrated luminosity is needed to observe the top signal over the background
at startup and the effects of a non-perfectly calibrated detector on its observability.\\
A study that uses a very simple selection in the  leptons + jets channel,
where $t\bar{t} \rightarrow W^+bW^-\bar{b}$ with a $W$ decaying hadronically and the other 
leptonically $ W\rightarrow e \nu_e (\mu\nu_{\mu})$,
 has been performed by the ATLAS collaboration~\cite{ivo}.
The selection requires 3 jets with transverse momentum $p_T>$ 40~GeV/c and one with 
$p_T>$~20 GeV/c, one isolated lepton with $p_T>$~20GeV/c and missing transverse energy 
$E_T>$~20 GeV/c$^2$.
In this selection the b-tagging information is deliberately not used since it might not be
optimized and calibrated in the initial phase of data taking.
The hadronic top is selected as the 3-jet combination with the highest 
transverse momentum: 2 out of the 3-jets would be resulting from a $W$ decay, therefore
only the combinations with a di-jet invariant mass  
$|m_{jj}-m_W|<10$ GeV/c$^2$ are kept.   
Figure 1 shows the expected distribution of the 3-jet invariant mass in a 100 
pb$^{-1}$ integrated luminosity sample.
\begin{figure}
\psfig{figure=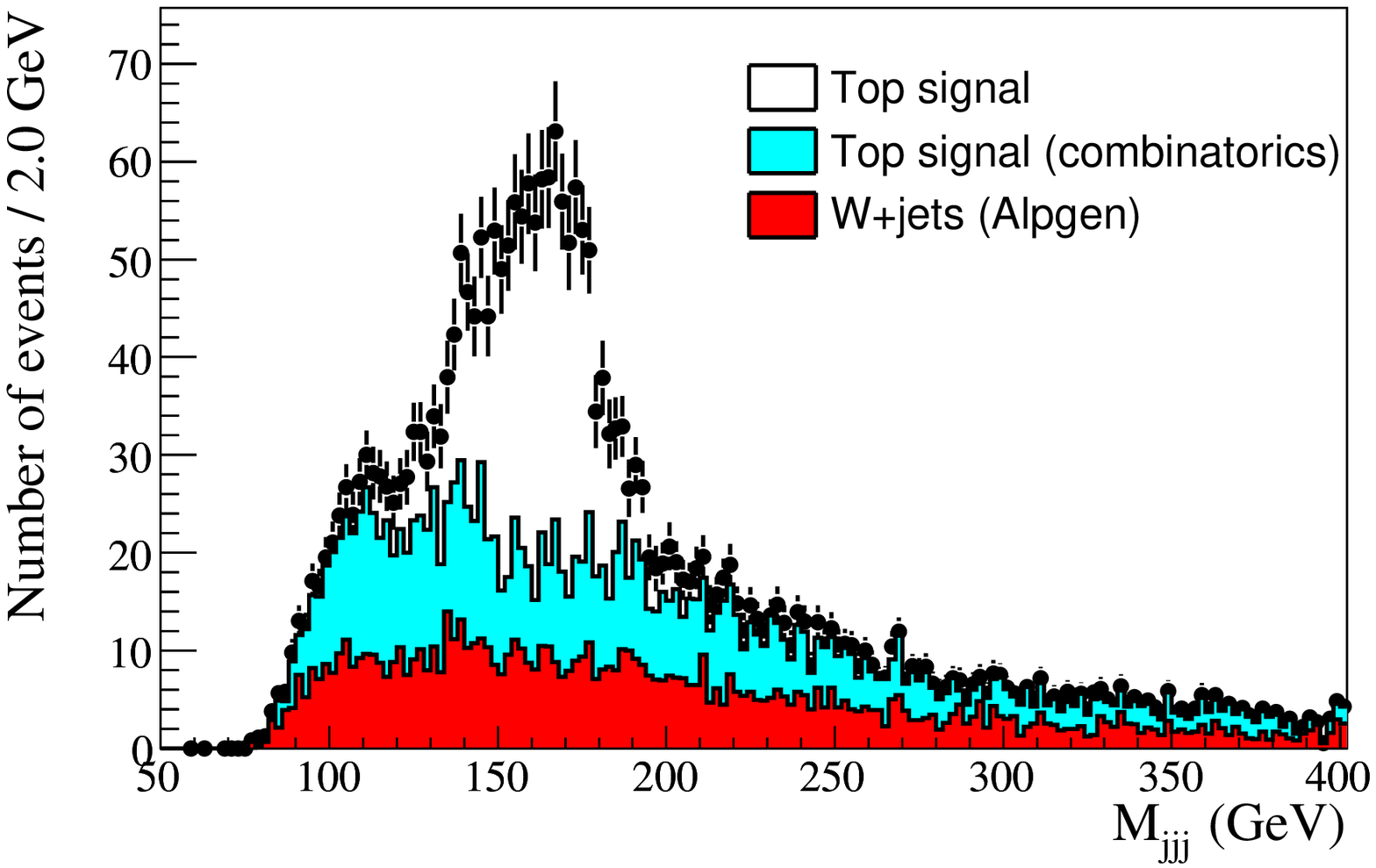,height=2 in}
\psfig{figure=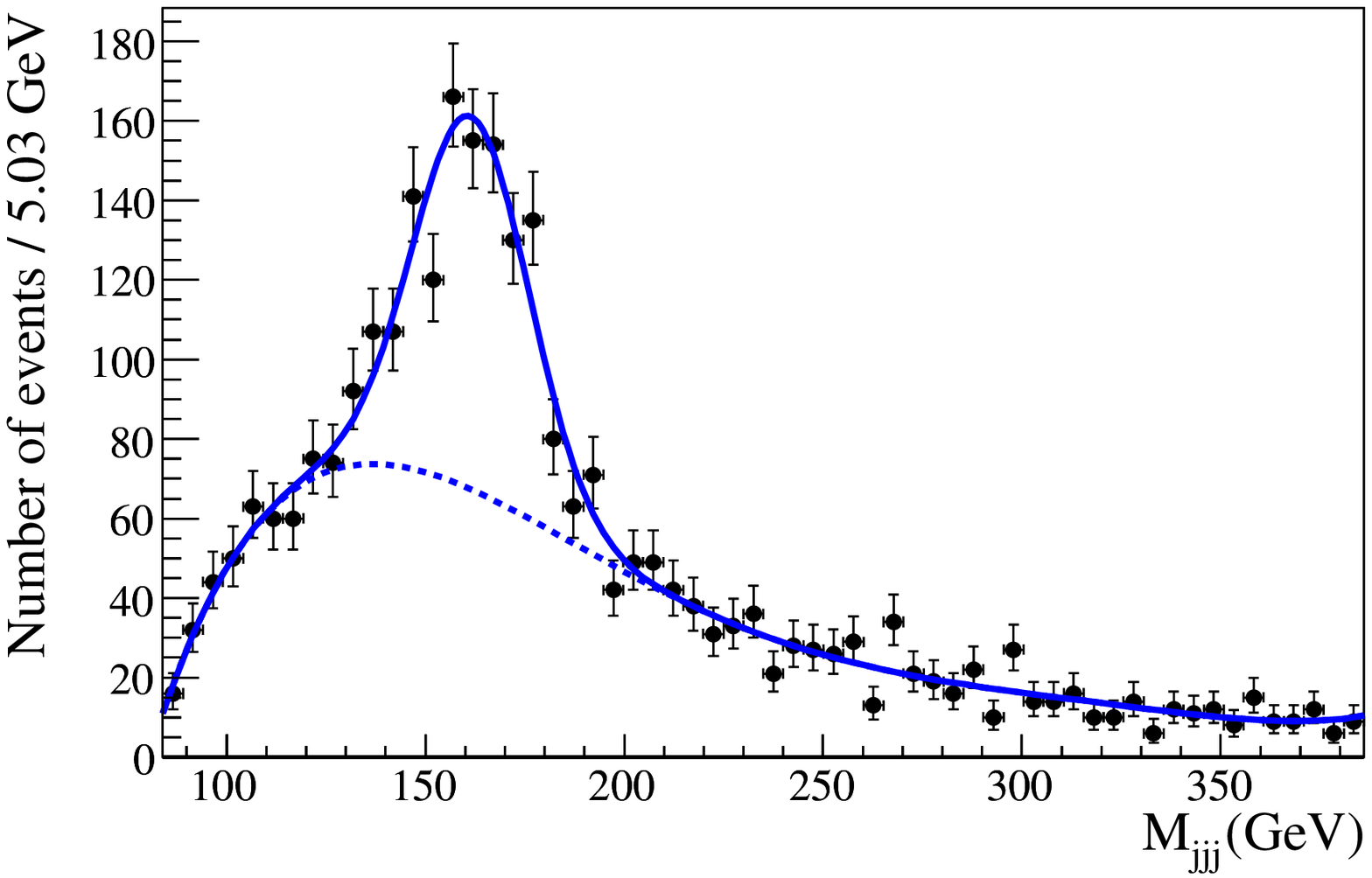,height=2in}
\caption{Expected distribution of the three-jet invariant mass after a cut on the di-jet system 
( left plot) and a fit ( right plot) in a 100 pb$^{-1}$ event sample. 
\label{fig:commtop}}
\end{figure}
The dominant background is the W+jets production  giving a contribution of the same
order as wrongly reconstructed $t\bar{t}$ events.
The signal over background ratio is about 0.7 and the relative statistical 
error is about 10\%. Overall the top cross-section could be determined with 
a total uncertainty of about 20\% with few hundred $pb^{-1}$ of integrated luminosity.

\section{Top cross-section evaluation}

In the leptons + jets  channel, a better accuracy on
the cross-section can be obtained by 
refining the selection and in particular by requiring 2 b-tagged jets.
To further reduce the background and 
combinatorics, a converging  kinematic fit to $m_W$ can be applied.
With 5 $fb^{-1}$ of integrated luminosity, a recent study by the CMS 
collaboration~\cite{art:CMSxsecsl} has extracted the 
$t\bar{t}$ cross-section with the following errors: $\delta \sigma/ \sigma = 0.6\%$ (statistical) 
$\pm9.2\%$ (systematical) $\pm 5.0\%$ (luminosity).\\
While the leptons+jet can be considered as the golden channel  since the 
background can be reduced by using simple cuts and the signal will be visible
very soon after start-up, promising results have been obtained also in the di-leptonic 
and fully hadronic channels, where both
$W$'s decay either leptonically (e,$\mu$) or hadronically, respectively.
A comparison of the performances in the different search channels,
as from recent studies by the 
CMS collaboration~\cite{art:CMSxsecsl,art:CMSxsechad}, can be read from Table~\ref{tab:CMSxsc}.

\begin{table}[t]
\label{tab:CMSxsc}
\caption{Breakdown of statistical, systematical and luminosity errors, main background sources,
efficiency and signal over background ratio S/B, for the cross-section studies in the 
lepton+jets, di-leptonic and hadronic channels. The S/B ratio for the lepton+jets channel 
doesn't take into account the background from $t\bar{t}$.}  
 
\vspace{0.4cm}
\begin{center}
\begin{tabular}{|c|c|c|c|c|c|c|c|c|}
\hline
 & syst $(\%)$ & stat $(\%)$ & lumi $(\%)$ & \mco{2}{|c|}{main syst. $(\%)$} & main bkg & eff & S/B \\
\hline
                           &     &     &    & b-tag   & 7   & $t\bar{t}$  &     &      \\
 10fb$^{-1}$               & 9.7 & 0.4 & 3  & PDF     & 3.4 & $W+j$ & 6.3 & 26.7  \\
 lepton+jets             &     &     &    & Pile-up & 3.2 &     &     &       \\ \hline
                           &     &     &    & PDF     & 5   &  $t\bar{t}$ with    &     &      \\
 10fb$^{-1}$               & 11  & 0.9 & 3  & b-tag   & 4   & $(W\rightarrow\tau \nu_{\tau}$ & 5 & 5.5  \\
 di-leptonic               &     &     &    & Jet E Scale (JES)     & 4   & and $\tau\rightarrow) l$     &     &       \\ \hline
 1fb$^{-1}$                &     &     &    & JES     & 11  &     &     &      \\
 hadronic                  & 20  &  3  & 5  & Pile-Up & 10  & QCD & 1.6 & 1/9  \\ \hline
\end{tabular}
\end{center}
\end{table}

\section{The top mass measurement}

In the lepton + jets  channel, after an event selection optimised  
not to bias the mass measurement, different methods have been exploited  
to extract the top mass ($m_t$).
The simplest is to perform a  fit to the invariant mass of the 3 jets arising 
from the hadronic top decay, but this suffers of the impact of poorly 
reconstructed jets  due to effects of FSR and to the semi-leptonic 
decay of b-quarks.
Another method, less affected by systematic errors, reconstructs event by event
the entire $t \bar t$ final state via a $\chi^2$ minimisation 
based on kinematic constraints: the energies of the leptons and jets, the jet directions and 
the 3 components of the reconstructed neutrino's are free to vary within their 
resolutions; $m_t$ is then fitted in slices of $\chi^2$ and is extrapolated
from a linear fit to the $m_t$ value corresponding to $\chi^2=0$~\cite{art:ATLmass}.
Alternatively, an event-by-event likelihood method 
which convolutes the resolution function of the event, or the so called ideogram, with 
the expected theoretical template can be used~\cite{art:CMSmass}.
A method which is appealing since it has independent systematic
errors, is to select high $p_T$ top pairs with $p_T> 200$ GeV/c:
in this case the 2 top quarks tend to be back to back and this can be 
used to reduce the backgrounds. Since the 3 jets on one hemisphere tend to overlap, 
the energy in a cone around the candidate top quark has to be collected making the measurement 
less sensitive to the jet energy calibration.
A summary of the different contributions to the error on $m_t$ for 
the different methods described above can be found in table~\ref{tab:mass}, as from an ATLAS 
study~\cite{art:ATLmass}.\\
\begin{table}[t]
\label{tab:mass}
\caption{Expected systematical and statistical error contributions  to the top mass
measurement expressed is $\delta m_t (GeV/c^2)$  for the 3 methods described in the text: the hadronic mass fit, the 
kinematic fit and the high $p_T$ selection.}  
\begin{center}
\begin{tabular}{|c|c|c|c|}
\hline
                         & had. top & kin. fit & high $p_T$ \\ \hline
 light jet E scale (1\%) & 0.2      &    0.2   &   - 		\\ \hline
 b-jet E scale (1\%)   	 & 0.7	&    0.7   &   -		\\ \hline
 b-quark fragmentation   & 0.1      &    0.1   &   0.3	\\ \hline
 ISR				 & 0.1      &    0.1   &   0.1 	\\ \hline
 FSR			   	 & 1.0	&    0.5   &   0.1	\\ \hline
 combinatorial bkg       & 0.1      &    0.1   &   -		\\ \hline
 mass rescaling		 & -	 	& 	-    &   0.9	\\ \hline
 Underlying event (10\%)	 & -	 	& 	-    &   1.3	\\ \hline
 total syst.		 & 1.3	&    0.9   &   1.6	\\ \hline
 stat. err. @10$fb^{-1}$  & 0.05     &    0.1   &   0.2      \\ \hline
\end{tabular}
\end{center}
\end{table}
As for the cross-section, $m_t$  can also be extracted from the di-leptonic
and the hadronic channels. The di-lepton channel has a clean signature, but 2
neutrino's need to be reconstructed, this can be done by applying a 
constrained fit assuming the $W$ mass and two equal masses for the 2 reconstructed top~\cite{art:CMSxsechad}.
With an integrated luminosity of 1 $fb^{-1}$, the statistical error on  $m_t$
would be of about 1.5 GeV/c$^2$ and the systematical about 4.2 GeV/c$^2$ .
In the hadronic channel a kinematic fit can be used to reconstruct both top quarks, but the 
measurement is affected by large QCD backgrounds~\cite{art:CMSxsechad}. 
With an integrated luminosity of  1 $fb^{-1}$ the statistical error would be of 
about 0.6 GeV/c$^2$ and the systematical about 4.2 GeV/c$^2$.

\section{Searches for new physics}

By reconstructing the top mass spectrum in $t\bar t $ leptons+jets events, 
resonances originated by the decay  
process  $p \bar p \rightarrow X \rightarrow t \bar t$ can be observed.
From preliminary studies by ATLAS a 1(2) $TeV/c^2$ mass $Z'$ boson 
produced with a cross-section of 4(3) $pb$  can be observed at about 3$\sigma$
significance with an integrated luminosity of about 5 $fb^{-1}$.\\
Already with 10 $fb^{-1}$ of data, flavour changing neutral currents, 
which are not allowed at tree level in the SM, can be observed with a 
sensitivity 2 orders of magnitude better than at Tevatron~\cite{art:ATLprop,art:CMSprop}.
Finally by studying the double differential angular distribution of $t \bar t$ decay products
and by comparing the observed values of the spin correlation observables and the SM
expectations,  the presence of 
anomalous couplings, Technicolor, spin 0/2 heavy resonances can be observed.
With an integrated luminosity of 10 $fb^{-1}$,  the  spin correlation 
observables can be extracted with a 3\% and 5\% statistical and systematical 
uncertainty, respectively~\cite{art:ATLas,art:CMSas}.

\section{Conclusions}

Top physics provides an excellent environment for calibrating  the detector and 
for testing  the SM predictions as well as new physics starting from the early days of 
data taking at the LHC.
A large effort has been made by the ATLAS and CMS Collaborations to be ready 
to analyse the top events from day one, by searching for better selection cuts, 
improving the generators and systematic errors understanding and exploring 
alternative analysis methods and decay channels.

\end{document}